\documentclass[aps,twocolumn,twoside,amsmath,bibnotes,floatfix]{revtex4}

\usepackage{epsfig}
\usepackage{natbib}

\begin{document}

\title{Flow equations for the one-dimensional Kondo lattice model:
Static and dynamic ground state properties}

\author{T. Sommer}
\affiliation{
  Institut f\"{u}r Theoretische Physik,
  Technische Universit\"{a}t Dresden, D-01062 Dresden, Germany
}

\date{\today}

\begin{abstract}
The one-dimensional Kondo lattice model is investigated by means of Wegner's
flow equation method. The renormalization procedure leads to an effective
Hamiltonian which describes a free one-dimensional electron gas and a
Heisenberg chain. The localised spins of the effective model are coupled by
the well-known RKKY interaction. They are treated within a Schwinger boson
mean field theory which permits the calculation of static and dynamic
correlation functions. In the regime of small interaction strength static
expectation values agree well with the expected Luttinger liquid
behaviour. The parameter $K_\rho$ of the Luttinger liquid theory is estimated
and compared to recent results from density matrix renormalization group
studies.
\end{abstract}

\maketitle

\section{Introduction}
\label{sec:introduction}

The fascinating subject of heavy fermion physics in rare-earth and actinide
systems has been a challenge for theoretical and experimental investigations
for decades \cite{fulde}. The intriguing properties of these materials are far
from being understood and still give us a lot of puzzles to solve. Theoretical
studies of the heavy fermion materials are based on several models like the
periodic Anderson model (PAM) \cite{hewson}. Another generic model is the
Kondo lattice model (KLM) which describes a noninteracting electron gas
coupled to localised spin moments via a Heisenberg spin interaction. The
Hamiltonian reads

\begin{equation}
{\cal H} = \sum_{k \sigma} \varepsilon_k \; c^\dagger_{k \sigma}
c_{k \sigma} + \frac{J}{2} \, \sum_{i \, \alpha \beta}
{\bf S}_i \, c^\dagger_{i \alpha} {\boldsymbol \sigma}_{\alpha \beta}
\, c_{i \beta},
\label{eq:kondo_hamiltonian}
\end{equation}
where $\varepsilon_k = - \sum_{ij} t_{ij} \, \text{e}^{ik(R_i-R_j)}$ is the
dispersion relation for the electrons on the lattice, $t_{ij}$ being the
hopping integrals. The parameter $J$ is the exchange integral of the local
spin interaction, the so called Kondo exchange.

We want to consider here the one-dimensional case which has been the subject
of numerous numerical and analytical investigations. Numerical studies were
based on the Quantum Monte Carlo (QMC) method \cite{troyer_wuertz}, exact
diagonalization (ED) studies \cite{tsunetsugu_rmp, tsunetsugu_prb}, the
density matrix renormalization group (DMRG) \cite{moukouri_caron,
  caprara_rosengren, shibata_jp, xavier_1, shibata_tsunetsugu} or the
numerical renormalization group (NRG) method \cite{yu_white}. Analytical
approaches comprised the bosonisation technique \cite{honner_gulacsi,
  mcculloch} or the renormalization group (RG) theory \cite{pivovarov_si}.

\begin{figure}[ht]
\centering
\epsfig{figure=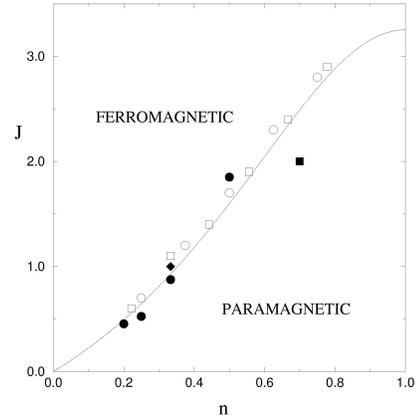,width=.75\columnwidth}
\caption{Phase diagram after \cite{honner_gulacsi}}
\label{fig:phase_diagram}
\end{figure}
The phase diagram of the one-dimensional KLM as a function of the Kondo
coupling $J$ and the band filling $n_c$ of the conduction electrons is quite
accurately known. In higher dimensions the KLM is believed to show the
well-known Doniach phase diagram \cite{doniach}. In contrast to the latter the
one-dimensional model does not exhibit a magnetically ordered phase in the
parameter regime of small interaction strengths $J$. In this parameter regime
the corresponding phase diagram is governed by a paramagnetic metallic phase
\cite{tsunetsugu_rmp}. There, the model is assumed to belong to the
universality class of the so called Luttinger liquids \cite{luttinger} which
possess gapless charge and spin excitations resulting in an algebraic decay of
the corresponding correlation functions. The asymptotic form for
density-density- and spin-spin-correlations are \cite{voit}
\begin{align}
\left< \delta n(x) \delta n(0) \right> = \frac{K_\rho}{(\pi \, x)^2}
&+ A_1 \cos(2 k_F x) \, x^{-1-K_\rho} \notag \allowdisplaybreaks \\
&+ A_2 \cos(4 k_F x) \, x^{-4 K_\rho} \allowdisplaybreaks \\[10pt]
\left< {\bf S}(x) \cdot {\bf S}(0) \right> = \frac{1}{(\pi \, x)^2}
&+ B_1 \cos(2 k_F x) \, x^{-1-K_\rho}.
\end{align}
The parameter $K_\rho$ is a model dependent constant which determines the
low-energy physics. Apart from the paramagnetic metallic phase for small $J/t$
the phase diagram further comprises a ferromagnetic ordered phase for large
$J$ and a spin liquid insulator phase at half-filling, $n_c = 1$. There are
two limiting cases in which the ground state has been proven to be
ferromagnetic \cite{tsunetsugu_rmp}. Firstly, the limit of vanishing electron
density $n_c \to 0$, secondly the case of infinite coupling strength $J/t \to
\infty$. The situation at half filling is special in the sense that it
exhibits finite gaps for spin and charge excitations at any finite coupling
$J$.

The KLM can be understood as an effective Hamiltonian of the above
mentioned PAM. It is connected to the PAM by a Schrieffer-Wolff
transformation \cite{schrieffer_wolff}. This property naturally raises
the question whether the localised spins in the KLM participate in the
formation of the Fermi surface, or in other words: Does $k_F = n_c
\pi/2$ or $k_F = (n_c+1) \pi/2$ hold? The size of the Fermi surface
can be read from the positions of singularities in certain correlation
functions. Recent results seemed to confirm the picture of a small
Fermi surface with $k_F = n_c \pi/2$ \cite{xavier_1}. However, a more
careful analysis which has recently been performed by Shibata
\emph{et al.} \cite{shibata_new} supports a large Fermi surface.

In this paper we shall apply the analytical method of continuous
unitary transformations (flow equations) proposed by Wegner
\cite{wegner} and G\l azek/Wilson \cite{glazek_wilson} to the
one-dimensional KLM. It was first applied to this model in arbitrary
dimensions by Stein \cite{stein}. He derived an analytical expression
for the RKKY interaction.

In Sec. \ref{sec:flow_equations} we shall give a short introduction
into the flow equation method. In Sec. \ref{sec:flow_klm} the method
will be applied to the one-dimensional KLM. By integrating out the
Kondo coupling between the conduction electrons and localised spins we
arrive at a decoupled system of a renormalized noninteracting
one-dimensional electron gas and a renormalized spin chain. In the
latter the spins interact via an effective spin exchange. Within
the framework of the flow equation method it is straightforward to
find expectation values and correlation functions, if the eigenvalue
problem of the effective model is known. In Sec. \ref{sec:results} we
shall show how the method can be used in order to verify the expected
characteristic behaviour of a Luttinger liquid. Previous
investigations of the one-dimensional KLM have mainly focused on static
properties like the momentum distribution or spin and charge
correlation functions. In this work we shall put special emphasis on the
investigation of dynamic properties and extend already existing
results for the dynamics.

\section{Flow equation method}
\label{sec:flow_equations}

To begin with we would like to sketch the concept of the flow equation
method which was independently developed by Wegner \cite{wegner} and
G\l azek/Wilson \cite{glazek_wilson} in 1994. Since then the method
has successfully been applied to a great number of problems,
\emph{e.g.} the electron-phonon-problem \cite{electron_phonon},
one-dimensional interacting fermion systems \cite{1d_systeme} or the
spin-boson-problem \cite{spin_boson}.

The basic idea of the flow equation method is the application of a
continuous set of unitary transformations to a given Hamiltonian
\begin{equation}
{\cal H}(l) = {\cal U}(l) \, {\cal H} \, {\cal U}^\dagger(l).
\end{equation}
Here $l$ means the continuous flow parameter. The purpose of this
procedure is that one wishes to diagonalize or at least simplify the
Hamiltonian. Thereby the parameters of the Hamiltonian become
renormalized. This treatment is translated into the language of
differential equations by using the expression
\begin{align}
\eta(l) = \frac{d \,{\cal U}(l)}{dl} \, {\cal U}^\dagger(l)
\end{align}
for the antihermitean generator $\eta(l) = -\eta^\dagger(l)$ of the
unitary transformation. The differential equation for the Hamiltonian
takes the simple form
\begin{equation}
\frac{d \, {\cal H}(l)}{d l} = \left[\eta(l) , {\cal H}(l)\right].
\label{eq:definition_flussgleichungen}
\end{equation}
The generator has to be suitably chosen. Wegner's approach starts from
a decomposition of the Hamiltonian into an unperturbed part ${\cal
H}_0$, whose eigenvalue problem is assumed to be known, and a
perturbation ${\cal H}_1$. Wegner's generator is given by
\begin{equation}
\eta(l) = \left[{\cal H}_0(l),{\cal H}(l)\right]
\label{eq:wegner_generator}
\end{equation}\\
which is simply the commutator between the unperturbed part ${\cal
H}_0(l)$ and the perturbation ${\cal H}_1(l)$. This generator
integrates out all interaction terms except for possible
degenerations \cite{wegner}. It finally leads to a diagonal or
block-diagonal effective Hamiltonian.

\section{Flow equations for the Kondo lattice model}
\label{sec:flow_klm}

We now turn to the derivation of the flow equations for the parameters
of the Hamiltonian. With this in mind we proceed as follows. Firstly, we give
the flow invariant Hamiltonian which includes new generated, effective
interactions. The flow invariant Hamiltonian then leads us to the
specification of the generator. Thereby we shall introduce some of the
necessary approximations within our approach.

\subsection{Flow equations for the Hamiltonian}
\label{subsec:hamiltonian_flow}

The first step in deriving the flow equations is the determination of
the generator $\eta(l)$. In a first step we wish to integrate out the Kondo
coupling between the conduction electrons and the localised spins, so the most
simple generator is
\begin{equation}
\eta(l) = \frac{1}{2N} \, \sum_{ikq \, \alpha \beta}
\eta^J_{kq}(l) \, {\bf S}_i \cdot {\boldsymbol \sigma}_{\alpha \beta}
\, c^\dagger_{k \alpha} c_{q \beta} \; \text{e}^{i (k-q) R_i} =:
\eta^J(l),
\label{eq:eta_0}
\end{equation}\\
where the coefficients $\eta^J_{kq}(l)$ are still unspecified. They
depend on the concrete choice of the generator. Wegner's approach
starts out from the generalised form of Eq. (\ref{eq:wegner_generator}).
If we take only the conduction electrons to be ${\cal H}_0$, we obtain
$\eta_{kq}(l) = (\varepsilon_k-\varepsilon_q) J_{kq}(l)$ for the
coefficients of the generator. The commutator between the generator
(\ref{eq:eta_0}) and the Hamiltonian (\ref{eq:kondo_hamiltonian})
gives rise to new, effective interactions. Using Wegner's approach
they enter the generator and are eventually integrated out. We shall
introduce a more general form of $\eta(l)$ below.

In order to see what kind of effective interactions emerge, let us
commute the initial Hamiltonian (\ref{eq:kondo_hamiltonian}) and the
generator of Eq. (\ref{eq:eta_0}). After some calculation we obtain
the following Hamiltonian
\begin{widetext}
\begin{align}
{\cal H}(l) &= \sum_{k\sigma} \varepsilon_k(l) :c^\dagger_{k\sigma}
c_{k\sigma}: + \frac{1}{2N} \, \sum_{kq} \chi_{kq}(l) \, :{\bf
S}_{k-q} \cdot {\bf S}_{q-k}: + E_c(l) 
\notag \allowdisplaybreaks \\
&+\frac{1}{2N} \, \sum_{kq \, \alpha \beta} J_{kq}(l) \, {\bf S}_{k-q}
\cdot {\boldsymbol \sigma}_{\alpha \beta} \, :c^\dagger_{k \alpha}
c_{q \beta}: + \frac{1}{4 N^2} \sum_{kpq \sigma} M_{kpq}(l) :{\bf
S}_{k-p} \cdot {\bf S}_{p-q}: \, :c^\dagger_{k \sigma} c_{q \sigma}:
\notag \allowdisplaybreaks \\
&+ \frac{1}{4 N^2} \sum_{kpq \alpha \beta} i D_{kpq}(l) \, ( {\bf
S}_{k-p} \times {\bf S}_{p-q} ) \cdot {\boldsymbol \sigma}_{\alpha
\beta} :c^\dagger_{k \alpha} c_{q \beta}:
\notag \allowdisplaybreaks \\[10pt]
&= {\cal H}_e (l) + {\cal H}_S (l) + E_c (l) + {\cal H}_J (l) + {\cal
H}_M (l) + {\cal H}_D(l),
\label{eq:flow_hamiltonian}
\end{align}
\end{widetext}
where $:{\cal X}:$ denote operators resulting from a decoupling scheme
which we shall discuss later.

Before we proceed let us take a closer look at equation
(\ref{eq:flow_hamiltonian}). The first line represents the block diagonal part
of the Hamiltonian since electron and spin operators are decoupled. It
contains a complicated RKKY-like spin interaction term between the local
moments. The second and third line comprise the nondiagonal or interaction
part. Aside from the Kondo coupling we get interactions between the local
moments which are either symmetric or antisymmetric with respect to
interchange of the sites. Correspondingly, the first one couples to the
electronic charge density, whereas the second one couples to the electronic
spin density. We restrict ourselves to these terms because they are the most
important ones in the regime of small interaction strength $J$. That way the
above Hamiltonian becomes flow invariant and Eq. (\ref{eq:flow_hamiltonian})
is valid for all flow parameters $l$. For $l = 0$ it represents the initial
Hamiltonian (\ref{eq:kondo_hamiltonian}). This implies the following initial
values of the parameters
\begin{align}
\varepsilon_k(l = 0) &= \varepsilon_k \, , \qquad J_{kq}(l = 0) = J
\notag \allowdisplaybreaks \\[10pt]
\chi_{kq}(l = 0) &= 0 \, , \qquad M_{kpq}(l = 0) = 0
\notag \allowdisplaybreaks \\[10pt]
D_{kpq}(l = 0) &= 0 \, , \qquad E_c(l = 0) = \sum_k \varepsilon_k
n_k.
\end{align}

The prefactor of any operator term of Eq. (\ref{eq:flow_hamiltonian})
controls the strength of the respective operator. Within the framework
of the flow equation method they are determined by corresponding
differential equations. With the choice of the generator $\eta(l)$
we can control which of these terms are kept and which are to be vanished.
Since the aim of our renormalization procedure is a blockdiagonal
Hamiltonian in which electron and spin operators are decoupled, we
have to remove all terms describing interactions between electron and
spin operators. The generator $\eta(l)$ of the continuous unitary
transformation has to be chosen appropriately.

With this in mind we can now write down the generator $\eta(l)$. Using
Wegner's approach we have to take into account the generated, effective
interactions. The generator reads
\begin{widetext}
\begin{align}
\eta(l) = \eta^J(l) + \eta^M(l) + \eta^D(l)
&= \frac{1}{2N} \sum_{kq \alpha \beta} \eta^J_{kq}(l) \; {\bf S}_{k-q}
\cdot {\boldsymbol \sigma}_{\alpha \beta} \, :c^\dagger_{k \alpha}
c_{q \beta}: + \frac{1}{4 N^2} \sum_{kpq \sigma} \eta^M_{kpq}(l) :{\bf
S}_{k-p} \cdot {\bf S}_{p-q}: \, :c^\dagger_{k \sigma} c_{q \sigma}:
\notag \\[10pt]
&+ \frac{1}{4 N^2} \sum_{kpq \alpha \beta} \eta^D_{kpq}(l) \; i ( {\bf
S}_{k-p} \times {\bf S}_{p-q} ) \cdot {\boldsymbol \sigma}_{\alpha
\beta} \, :c^\dagger_{k \alpha} c_{q \beta}:
\label{eq:generator}
\end{align}
\end{widetext}
\noindent
and the prefactors $\eta^J_{kq}(l)$, $\eta^D_{kpq}(l)$ and
$\eta^M_{kpq}(l)$ are determined by Eq. (\ref{eq:wegner_generator}).

After the transformation, \emph{i.e.} in the limit $l \to \infty$,
only the first line of Eq. (\ref{eq:flow_hamiltonian}) remains. It
represents the diagonal part ${\cal H}_0(l)$. This effective
Hamiltonian can be used to easily calculate physical properties.
The nondiagonal part ${\cal H}_1(l)$ vanishes for $l \to \infty$ and the
effective Hamiltonian $\tilde{\cal H} := {\cal H}(l = \infty)$ then reads
\begin{align}
\tilde{\cal H} &= \sum_{k\sigma}
\tilde{\varepsilon}_k :c^\dagger_{k\sigma} c_{k\sigma}:
+ \frac{1}{2N} \, \sum_{kq} \tilde{\chi}_{kq} \, :{\bf S}_{k-q} \cdot
{\bf S}_{q-k}: + \tilde{E}_c
\notag \\ 
&= \tilde{\cal H}_e + \tilde{\cal H}_S + \tilde{E}_c.
\label{eq:eff_modell}
\end{align}
In the following we shall denote all renormalized variables by a
tilde. As Eq. (\ref{eq:eff_modell}) tells us the effective model
will consist of a one-dimensional noninteracting electron gas and a
Heisenberg spin chain with renormalized parameters.

We now have all ingredients needed to derive the flow equations for
the parameters of the Hamiltonian. Before doing this we want to look
at the approximations that have to be done. Firstly, we neglect
interactions of order ${\cal O}(J^3)$ and higher in the Hamiltonian
(\ref{eq:flow_hamiltonian}). Secondly, we decouple higher operator
products in order to reduce them to those appearing in ${\cal H}(l)$.
This gives rise to operator expressions of the form $:{\cal X}:$.
They refer to fluctuation operators and mean either a normal
order product of fermionic operators or a Hartree-Fock-decoupling
scheme of spin operator products
\begin{align}
:c^\dagger_{k \sigma} c_{k \sigma}: &=  c^\dagger_{k \sigma} c_{k
\sigma} - \langle c^\dagger_{k \sigma} c_{k \sigma} \rangle,
\allowdisplaybreaks \\[10pt]
:{\bf S}_{k-q} \cdot {\bf S}_{q-k}: &= {\bf S}_{k-q} \cdot {\bf
S}_{q-k} - \langle {\bf S}_{k-q} \cdot {\bf S}_{q-k} \rangle.
\end{align}
The thermodynamic average will here be taken with respect to the effective
model $\tilde{\cal H}$, Eq. (\ref{eq:eff_modell}), which describes a decoupled
system of a simple Fermi gas (electrons) and a Heisenberg spin chain with
long-range interactions. These expectation values are therefore
$l$-independent. The decoupling leads to a formal temperature dependence of
the flow equations. Here we consider only the ground state properties,
\emph{i.e.} $T = 0$. For the sake of simplicity we introduce the abbreviation
$S(k-q) := \langle {\bf S}_{k-q} \cdot {\bf S}_{q-k} \rangle$ for the spin
correlation function. One may also think of other expectation values like
$\langle {\bf S}_{k-q} \times {\bf S}_{q-k} \rangle$ or $\langle {\bf S}_q
\rangle$. Since we consider the limit of small Kondo coupling $J$, the
system is in the paramagnetic phase, where no symmetry is broken. Therefore
these expectation values vanish.

Evaluating the commutator between the generator (\ref{eq:generator}) and the
Hamiltonian (\ref{eq:flow_hamiltonian}) we arrive at the flow equations for
the parameters of the Hamiltonian. For the sake of clarity the $l$-dependence
of all parameters is dropped. The electronic single particle energies
$\varepsilon_k$ obey the following differential equation
\begin{equation}
\frac{d \varepsilon_k}{d l} = \frac{1}{2 N} \sum_q S(k-q) \,
\eta^J_{kq} \, J_{qk}.
\label{eq:dgl_ek}
\end{equation}
Here $S(k-q)$ is the local moment's spin correlation function which has to be
evaluated with respect to the renormalized Hamiltonian $\tilde{\cal H}$. It is
therefore $l$-independent. As the effective model is not known before the end
of the transformation we have to solve all flow equations self-consistently.

For the paramter $\chi_{kq}$ of the effective spin interaction we obtain the
following flow equation
\begin{align}
\frac{d \chi_{kq}}{d l} &= (n_k - n_q) \, \eta^J_{kq} \, J_{qk}.
\label{eq:dgl_rkky}
\end{align}
The occupation numbers $n_k$ which enter the above equation are again formed
with respect to the effective model. The constant $E_c$ of ${\cal H}_0(l)$
follows
\begin{align}
\frac{d E_c}{d l} &= \frac{1}{N} \sum_{kq} (n_k - n_q) \, S(k-q) \,
\eta^J_{kq} \, J_{qk}.
\label{eq:dgl_const}
\end{align}
We restrict the renormalization of the effective interaction terms to
contributions of order ${\cal O}(J^2)$. Therefore both coupling parameters
$D_{kpq}$ and $M_{kpq}$ obey the same flow equation
\begin{align}
\frac{d D_{kpq}}{d l} &= \frac{1}{2} (\eta^J_{kp} J_{pq} + \eta^J_{qp}
J_{kp}) -(\varepsilon_k - \varepsilon_q) \, \eta^D_{kpq}.
\end{align}
The first term is responsible for the generation of the effective coupling
while the second contribution, which is always negative, ensures the vanishing
at the end of the renormalization procedure. Finally for the flow equation of
the Kondo coupling we find
\begin{align}
\frac{d J_{kq}}{d l} &= -(\varepsilon_k - \varepsilon_q) \eta^J_{kq}
\notag \allowdisplaybreaks \\
&+ \frac{1}{N} \sum_p \left( n_p - \frac{1}{2} \right) (\eta^J_{kp}
J_{pq} + \eta^J_{qp} J_{pk})
\notag \allowdisplaybreaks \\
&+ \frac{3}{8N} \sum_p (\eta^J_{kp} D_{pkq} + \eta^J_{qp} D_{pqk})
\notag \allowdisplaybreaks \\
&+ \frac{3}{8N} \sum_p (\eta^D_{kqp} J_{pk} + \eta^D_{qkp} J_{pk})
\notag \allowdisplaybreaks \\
&- \frac{1}{8N} \sum_p (\eta^J_{kp} D_{p,p+q-k,q} + \eta^J_{qp}
D_{p,p+k-q,k})
\notag \allowdisplaybreaks \\
&- \frac{1}{8N} \sum_p (\eta^D_{k,p+q-k,p} J_{pk} +
\eta^D_{q,p+k-q,p} J_{pk}),
\label{eq:dgl_kondo}
\end{align}
where we have taken into account correction terms up to order
${\cal O}(J^3)$. Therefore we expect to find reasonable results only in the
parameter regime of small coupling strength $J/t$. As this ratio increases
further correction terms have to be included. The flow equations
(\ref{eq:dgl_ek}) to (\ref{eq:dgl_kondo}) represent a closed system of first
order differential equations, whose solution can only be found by numerical
integration.

\subsection{Approximations for the effective model}
\label{subsec:sbmft}

In the preceeding subsection we have derived flow equations for the parameters
of the Hamiltonian. As to solve them we still need an analytical expression
for the spin correlation function $S(k-q)$. As it describes spin correlations
of the effective model, we are dealing here with a one-dimensional Heisenberg
chain with long-range interactions whose exact solution is not known. Hence,
we have to resort to further approximations. We stress here that this is the
most crucial approximation within our approach because it strongly affects all
renormalized quantities. Since the spin interaction is the result of the
continuous unitary transformation it is not known until the transformation is
completely performed.\\
As our approach is only valid for small $J/t$, \emph{i.e.} for the
paramagnetic metallic phase with no broken symmetry, we use the Schwinger
boson formalism to describe the spin system \cite{arovas_auerbach}. It
preserves the rotational invariance of the spin Hamiltonian. The spin
operators are expressed in terms of Schwinger bosons $a_{i \sigma}$ and
$a^\dagger_{i \sigma}$ according to
\begin{equation}
S^\gamma_i = \frac{1}{2} \; \sum_{\sigma \sigma'} a^\dagger_{i \sigma}
\, \sigma^\gamma_{\sigma \sigma'} a_{i \sigma'},
\end{equation}
where $\sigma^\gamma_{\sigma \sigma'}$ stands for the Pauli spin matrix. Since
the occupation number for bosons is not restricted, a local constrained of the
form $\sum_\sigma \; a^\dagger_{i \sigma} a_{i \sigma} = 2 S$ must be
enforced.

We follow here the procedure of Trumper \emph{et al.}
\cite{trumper_manuel_gazza_ceccatto} or of Ceccatto \emph{et al.}
\cite{ceccatto_gazza_trumper} and introduce two fields
\begin{align}
A_{ij} &= \frac{1}{2} \; \sum_\sigma \sigma \, a_{i \sigma} a_{j
\bar{\sigma}} = -A_{ji}\\
\intertext{and}
B_{ij} &= \frac{1}{2} \; \sum_\sigma a^\dagger_{i \sigma} a_{j \sigma}
= B^\dagger_{ji}
\end{align}
describing antiferro- and ferromagnetic correlations, respectively
($\bar{\sigma} = -\sigma$). This yields to the following Hamiltonian
\begin{equation}
\tilde{\cal H}_S = \sum_{ij} \; J_{ij} \, {\cal N} (B^\dagger_{ij}
B_{ij}) - A^\dagger_{ij} A_{ij}.
\end{equation}
The expression ${\cal N} ({\cal O})$ stands for a normal order product of
boson operators. The Hamiltonian is now biquadratic with respect to the
Schwinger boson operators. We use a mean field theory in order to decouple the
biquadratic terms. By using the mean field parameters $\langle B_{ij} \rangle$
and $\langle A_{ij} \rangle$ and replacing the local constrained by a global
one we obtain a Hamiltonian which can easily be diagonalized via a Bogolubov
transformation. Introducing new boson operators $\alpha_{k \sigma} = u_k \,
a_{k \sigma} + i \sigma v_k \, a^\dagger_{-k \bar{\sigma}}$ we obtain
\begin{equation}
\tilde{\cal H}_S = \sum_{q \sigma} \, \omega_q \, \alpha^\dagger_{q
\sigma} \alpha_{q \sigma} + \frac{1}{2} \sum_{q \sigma} \omega_q,
\label{eq:sbmf_hamiltonian_operatorform_diagonal}
\end{equation}
with $\omega(q) = \sqrt{(\gamma_B(q)-\lambda)^2-\gamma^2_A(q)}$ representing
the energies of the elementary excitations $\alpha^\dagger_{q \sigma}$ of the
spin system. Here the quantities $\gamma_A(q) = \frac{i}{2} \, \sum_{R_{ij}}
J_{ij} \langle A_{ij} \rangle \, \text{e}^{i q R_{ij}}$ and $\gamma_B(q) =
\frac{1}{2} \sum_{R_{ij}} J_{ij} \langle B_{ij} \rangle \, \text{e}^{i q
  R_{ij}}$ are used. The mean field parameters $\langle B_{ij} \rangle$ and
$\langle A_{ij} \rangle$ and the Lagrange parameter $\lambda$ have to be
determined selfconsistently by solving the corresponding saddle point
equations.

Finally we find an analytic expression for the spin correlation function which
for $T = 0$ reads
\begin{equation}
S(q)_{T = 0} = \frac{1}{4 N} \sum_k \left( \cosh\left[ 2(\theta_k -
\theta_{k+q}) \right] - 1 \right),
\label{eq:sq_0}
\end{equation}
with $\theta_k = - \frac{1}{2} \tanh^{-1} \left(
\frac{\gamma_A(k)}{\gamma_B(k)-\lambda} \right)$.\\

Compared to methods like the Bethe ansatz for the nearest-neighbour Heisenberg
chain the approximative Schwinger boson treatment discussed above has the
advantage that as many interaction terms as possible can be taken into
account. With the approximation for the effective model we are able to
describe the one-dimensional KLM consistently within the framework of the flow
equation method. Any physical quantity we are interested in can be evaluated
within the present approach. Especially, we emphasise that nothing has to be
put in by hand.

\subsection{Expectation values and correlation functions}
\label{subsec:correlation_functions}

We now turn to the calculation of expectation values and correlation
functions. In this subsection we give the essentials for the derivation of
certain important expectation values and correlation functions. We shall
discuss the results in Sec. \ref{sec:results}.

The retarded Green's function between operators $A$ and $B$ is
in general defined as the following commutator or anticommutator
relation
\begin{equation}
G_{AB}(t) = -i \theta(t) \langle \langle A(t); B \rangle \rangle =
-i \theta(t) \langle [A(t), B]_\pm \rangle,
\end{equation}
depending on the statistics under consideration. The thermodynamic
average and the time-dependence have to be taken with respect to the
full Hamiltonian. One can exploit the invariance of the trace under
unitary transformations and obtains
\begin{equation}
G_{AB}(t) = -i \theta(t) \langle \langle \tilde{A}(t); \tilde{B}
\rangle \rangle_{\tilde{\cal H}}.
\end{equation}
Now the thermodynamic average and the time-dependence are taken
with respect to the effective model. According to the transformation
of the Hamiltonian we also have to transform the operators. They
obey a similar flow equation as the Hamiltonian
\begin{equation}
\frac{dA(l)}{dl} = \left[\eta(l), A(l) \right].
\label{eq:operatortrafo}
\end{equation}

The commutation between $\eta(l)$ of Eq. (\ref{eq:generator}) and the electron
operator $c_{k \sigma}$ leads to the following operator structure
\begin{align}
c_{k \sigma} (l) = \alpha_k (l) \, c_{k \sigma}
&+ \frac{1}{N} \sum_{q} \sigma \, \gamma_{kq} (l) \,
S^z_{k-q} c_{q \sigma} \notag \\
&+ \frac{1}{N} \sum_{q} \gamma_{kq} (l) \,
S^{-\sigma}_{k-q} c_{q \bar{\sigma}},
\label{eq:operator_e}
\end{align}
where we have taken only the correction terms into account that couple to one
local moment. The initial conditions of the parameters are $\alpha_k(l = 0) =
1$ and $\gamma_{kq}(l = 0) = 0$. We transform the spin operator according to
\begin{align}
S^z_i (l) = \beta (l) S^z_i &+ \frac{1}{N} \sum_{kq
\sigma} \, \zeta_{kq} (l) \; \sigma S^\sigma_i
\text{e}^{i(k-q)R_i} \, c^\dagger_{k \bar{\sigma}} c_{q \sigma}
\label{eq:operator_sz_l} \\
S^\sigma_i (l) = \beta (l) S^\sigma_i &+
\frac{1}{N} \sum_{kq \sigma'} \, \zeta_{kq} (l) \; \sigma
S^\sigma_i \, \text{e}^{i(k-q)R_i} \, \sigma' \, c^\dagger_{k \sigma'} c_{q
\sigma'} \notag \\
&+ \frac{2}{N} \sum_{kq} \zeta_{kq} (l) \; \sigma \, S^z_i \,
\text{e}^{i(k-q)R_i} \, c^\dagger_{k \sigma} c_{q \bar{\sigma}}.
\label{eq:operator_ssigma_l}
\end{align}
Here the initial parameters are $\beta(l = 0) = 1$ and $\zeta_{kq} (l = 0) =
0$.

In order to derive the flow equations for the parameters of the operator
transformations we have to use an equivalent decoupling scheme as for the
Hamiltonian. We finally arrive at the following differential equations
\begin{align}
\frac{d \alpha_k}{dl} &= \frac{1}{2 N} \sum_q S(k-q) \, \eta^J_{kq}
\gamma_{qk}
\label{eq:flussgleichung_alpha} \\[10pt]
\frac{d \gamma_{kq}}{dl} &= \frac{1}{2} \, \eta^J_{qk} \, \alpha_{k}
\label{eq:flussgleichung_gamma}
\end{align}
for the parameters of the electron operator transformation and
\begin{align}
\frac{d \beta}{dl} &= -\frac{2}{N^2} \, \sum_{kq} \eta^J_{kq} \,
\zeta_{kq} \, n_k ( 1 - n_q )
\label{eq:flussgleichung_beta} \\[10pt]
\frac{d \zeta_{kq}}{dl} &= \frac{1}{2} \, \beta \, \eta^J_{qk}
\label{eq:flussgleichung_zeta}
\end{align}
for the parameters of the spin operator transformations. We notice that the
spin correlation function $S(k-q)$ of the effective model enters the flow
equation of $\alpha_k(l)$ whereas the occupation numbers $n_k$ govern the flow
equation of $\beta(l)$. We restrict the flow equations for the correction
terms to first order contributions in the Kondo coupling. Going beyond this
approximation could bring us up against the violation of certain summation
rules which have to be fullfilled. We can combine the above equations to
obtain flow invariant expressions. The expectation values $S(k-q)$ and $n_k$
are taken with respect to the effective Hamiltonian and are therefore
$l$-independent. We arrive at
\begin{equation}
\alpha^2_k(l) + \frac{1}{N}\sum_{q} S(k-q) \gamma^2_{kq}(l) = 1
\label{eq:const_el}
\end{equation}
and
\begin{equation}
\beta^2 (l) + \frac{4}{N^2} \sum_{kq} \zeta^2_{kq} (l) \,
n_k ( 1 - n_q ) = 1,
\label{eq:const_spin}
\end{equation}
which displays the unitarity of the transformation.

After determining the operator transformation we are now able to calculate
static and dynamic correlation functions that characterise the ground state
properties of the one-dimensional KLM. One of the most important quantities is
the momentum distribution $n(k)$ which reads
\begin{align}
n(k) = \langle c^\dagger_{k\sigma} c_{k\sigma} \rangle =
\tilde{\alpha}^2_k \, n_k + \frac{1}{N} \sum_q \,
\tilde{\gamma}^2_{kq} \, S(k-q) \, n_q.
\label{eq:nk}
\end{align}
For a Luttinger liquid we expect a continuous behaviour with respect to the
momentum $k$ and a power law singularity at the Fermi momentum. The position
of this singularity fixes the size of the Fermi surface.

The static correlation function of the local moments $S_{ff}(q)$ indicates the
phase transition from the paramagnetic phase into the ferromagnetic phase on
increasing the Kondo coupling $J$. Within our approach it is given by
\begin{align}
S_{ff}(q) &= \langle {\bf S}_q \cdot {\bf S}_{-q} \rangle \notag \\
&= \tilde{\beta}^2 \, S(q) + \frac{4}{N^2} \, \sum_{kp}
\tilde{\zeta}^2_{kp} \, S(k-p+q) \, n_k (1-n_p).
\label{eq:Sff}
\end{align}
We can also evaluate the static charge correlation function $C(q)$ and the
static spin correlation function $S_{cc}(q)$ of the electrons. Their rather
lengthy expressions are given in the appendix.

The flow equation formalism allows us to calculate dynamic quantities. The
first quantities we look at are the one-particle spectral functions $A_\pm (k,
\omega)$ of the conduction electrons which measure occupied and empty states
of the conduction electrons.
\begin{align}
A_+ (k, \omega) &= \int_{-\infty}^{\infty} dt \langle c_{k \sigma} (t)
\, c^\dagger_{k \sigma} \rangle \, \text{e}^{i \omega t}
\notag \\[10pt]
&= \tilde{\alpha}^2_k \, (1-n_k) \, \delta(\omega -
\tilde{\varepsilon}_k)
\notag \\
&+ \frac{1}{2 N^2} \, \sum_{qp} \, \tilde{\gamma}^2_{kq} \, ( u_p \,
v_{k+p-q} - v_p \, u_{k+p-q} )^2 \notag \\
&\hspace{1.75cm} \times (1-n_q) \, \delta( \omega -
\tilde{\varepsilon}_q - \tilde{\omega}_p - \tilde{\omega}_{k+p-q})
\allowdisplaybreaks \\[10pt]
A_- (k, \omega) &= \int_{-\infty}^{\infty} dt \langle c^\dagger_{k
\sigma} (t) \, c_{k \sigma} \rangle \, \text{e}^{i \omega t}
\notag \\[10pt]
&= \tilde{\alpha}^2_k \, n_k \, \delta(\omega - \tilde{\varepsilon}_k)
\notag \\
&+ \frac{1}{2 N^2} \, \sum_{qp} \, \tilde{\gamma}^2_{kq} \, ( u_p \,
v_{k+p-q} - v_p \, u_{k+p-q} )^2 \notag \\
&\hspace{2.5cm} \times n_q \, \delta( \omega - \tilde{\varepsilon}_q +
\tilde{\omega}_p + \tilde{\omega}_{k+p-q}).
\end{align}
Here $u_k$ and $v_k$ are the coefficients of the Bogolubov transformation
used to diagonalize the Schwinger boson mean field Hamiltonian
(\ref{eq:sbmf_hamiltonian_operatorform_diagonal}). The spectral functions
$A_\pm (k, \omega)$ comprise two contributions. The first term ($\sim
\tilde{\alpha}^2$) represents a coherent quasiparticle excitation. The second
term is an incoherent background. It is important to note that the elementary
excitations of the spin system of the effective Hamiltonian $\tilde{\omega}_q$
enter the latter contribution. The electronic density of states defined by
\begin{equation}
\rho(\omega) = - \frac{1}{N} \sum_k \; \frac{1}{\pi} \; \text{Im} \,
G(k,\omega),
\label{eq:rho}
\end{equation}
with $G(k,\omega)$ being the electronic Green's function, can also be
calculated.\\

Another important quantity is the dynamic spin structure factor
$S_{ff}(q,\omega)$ of the local moments
\begin{widetext}
\begin{align}
S_{ff}(q,\omega) = \int_{-\infty}^{\infty} dt \, \langle {\bf S}_q (t)
\cdot {\bf S}_{-q} \rangle \, \text{e}^{i \omega t} &= \frac{1}{2 N}
\, \sum_{p} \, \tilde{\beta}^2 \, ( u_p \, v_{p+q} - v_p \, u_{p+q}
)^2 \, \delta(\omega - \tilde{\omega}_p - \tilde{\omega}_{p+q})
\allowdisplaybreaks \notag \\
&+ \frac{2}{N^3} \sum_{kpp'} \tilde{\zeta}^2_{kp} (u_{p'} v_{p'+k-p+q}
- v_{p'} u_{p'+k-p+q} )^2 \, n_k \, (1-n_p) \notag \\
& \hspace{4cm} \times \delta(\omega - \tilde{\omega}_{p'} -
\tilde{\omega}_{p'+k-p+q} + \tilde{\varepsilon}_k -
\tilde{\varepsilon}_p).
\label{eq:S_qw}
\end{align}
\end{widetext}
The first line describes only the spin excitations of the effective model in
terms of Schwinger bosons. The second line of Eq. (\ref{eq:S_qw}) results from
the coupling of the local moments to electronic particle-hole excitations of
the effective Hamiltonian. In addition, we can also calculate the dynamic spin
structur factor of the conduction electrons $S_{cc}(q,\omega)$ which is found
in the appendix.

\section{Results}
\label{sec:results}

After having derived theoretical expressions for various correlation functions
from the flow equation approach we now turn to present the outcome of the
numerical solution of the flow equations (\ref{eq:dgl_ek}) -
(\ref{eq:dgl_kondo}) and (\ref{eq:flussgleichung_alpha}) -
(\ref{eq:flussgleichung_zeta}) . We start with the result for the parameters
of the Hamiltonian and subsequently show our findings for static and dynamic
correlation functions. We shall show to what extend the statics reflects the
expected Luttinger liquid behaviour. We also clarify the possibility of the
approach to describe the quantum phase transition on increasing coupling
strength.

\subsection{Parameters of the Hamiltonian}
\label{subsec:parameters}

In order to solve Eqs. (\ref{eq:dgl_ek}) - (\ref{eq:dgl_kondo}) we used a
Runge Kutta algorithm. The complexity of the differential equation restricted
our system size to $N = 120$. Remember that the spin correlation function
$S(k-q)$ which enters the flow equations has to be calculated with respect to
the effective model (\ref{eq:eff_modell}). Therefore the parameters of
the Hamiltonian had to be determined self-consistently.

\begin{figure}[ht]
\centering
\epsfig{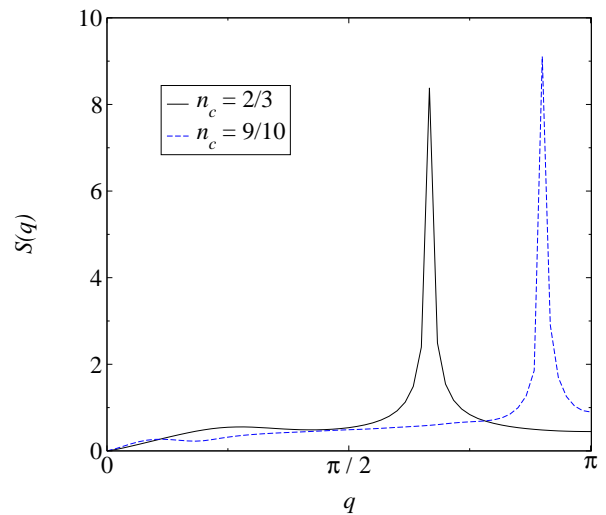}
\caption{Spin correlation function $S(q)$ of the effective model for two
  different fillings $n_c$.}
\label{fig:Sq}
\end{figure}
The spin correlation function $S(q)$ of the effective model plays an important
role. We therefore start our discussion with $S(q)$ which is shown in
Fig. \ref{fig:Sq}. The main feature is the dominant peak that shows up exactly
at the wave vector $q = 2 k_F^c = n_c \pi$, where $k_F^c$ is the Fermi
momentum of the conduction electrons. As we shall see later the pronounced
structure has severe consequences for other quantities that are related to
$S(q)$. The pronounced peak is due to the special excitation spectrum of the
Schwinger bosons. The other main property of $S(q)$ is the vanishing
ferromagnetic component ($q = 0$) which can easily be understood from
Eq. (\ref{eq:sq_0}).

\begin{figure}[ht]
\centering
\epsfig{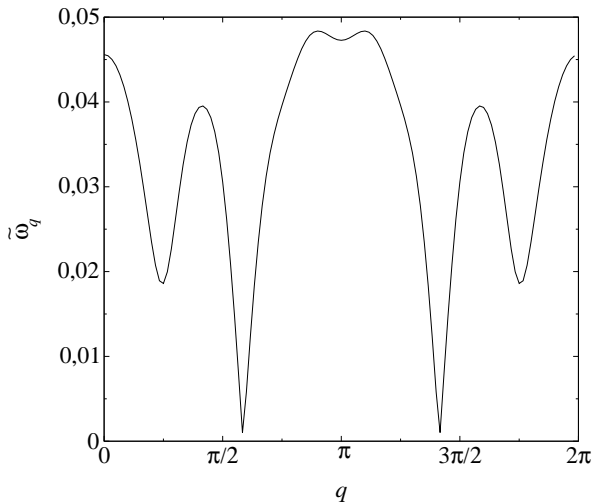}
\caption{Elementary excitations $\tilde{\omega}_q$ at $n_c = 2/3$ and
$J/t = 1.5$.}
\label{fig:wq}
\end{figure}
The elementary excitations $\tilde{\omega}_q$ of the spin system of
$\tilde{\cal H}$ are shown in Fig. \ref{fig:wq}. They exhibit a small but
finite gap at $q = k_F^c \mod \pi$. This small gap is responsible for the
strong peak in $S(q)$. It manifests the rotational invariance of the ground
states and is an artifact of the Schwinger boson approach as we are dealing
here with half integer spins ($S = 1/2$) which may have a gapless excitation
spectrum. Nevertheless, within the Schwinger boson approach a vanishing gap
would give rise to a ground state with broken symmetry that contradicts the
assumption of a rotational invariant paramagnetic phase. However, the
important point is the position of this gap. It determines the maximum of the
spin correlation function which turns out to be at the expected
position. Therefore we assume that a description in terms of spinons should
not change these results decisively. Looking at Eq. (\ref{eq:sq_0}) we see
that always pairs of excitations enter the equation for $S(q)$ so that the
maximum of the spin correlation function is found at $q = 2 k_F^c = n_c
\pi$.

At this point we add that we found solutions for the saddle point equations of
the SBMFT only in the parameter regime $1/2 < n_c < 1$. The case of half
filling is special in the sense that there exists a gapped spin liquid
phase. It remains an open question whether the present approach can also be
used to describe this phase. Below $n_c = 1/2$ the dominance of the
ferromagnetic components in the RKKY coupling ${\cal J}_{ij}$ prevents a
solution of the saddle point equations of the SBMFT.

\begin{figure}[ht]
\centering
\epsfig{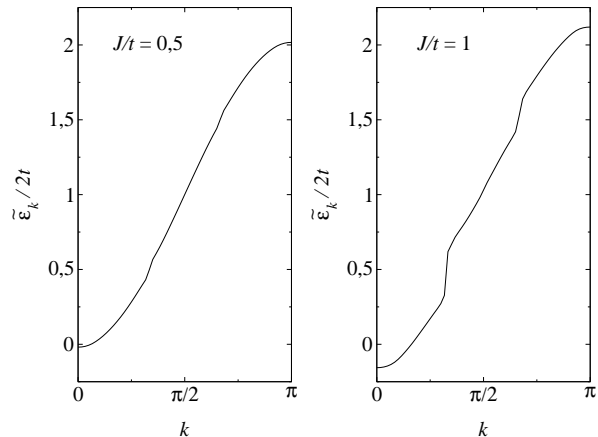}
\caption{Renormalized electronic single particle energies
$\tilde{\varepsilon}_k$ at $n_c = 2/3$ and different values $J/t$.}
\label{fig:ek}
\end{figure}
Finally we discuss the renormalized electronic single-particle energies. The
dispersion relation $\tilde{\varepsilon}_k$ is presented in
Fig. \ref{fig:ek}. We assume the unrenormalized single-particle energies to
follow a tight binding dispersion $\varepsilon_k = -2t(\cos k-1)$, where we
set the bottom of the band equal to zero and $t = 1/2$. We recognise two basic
features for $\tilde{\varepsilon}_k$. The first is a broadening of the
band. The effective band width is enlarged compared with the original band
width $W = 4t$. The other one is a decreasing density of states at the Fermi
momentum $k = k_F^c$ and at $k = \pi-k_F^c$. This property is mainly due to
the dominant peak structure in the spin correlation function $S(q)$ at $q = 2
k_F^c$. The wave vector $q = 2 k_F^c$ connects the two points of the Fermi
surface. Therefore the energies near the Fermi surface become more strongly
renormalized than energies near the band edge. The pseudo-gap like structure
at $k = k_F^c$ is thus due to the strong spin fluctuations at $q = 2
k_F^c$. As we are going to see this behaviour shall have consequences for the
electronic density of states $\rho(\omega)$. As the Luttinger liquid theory
expects $\rho(\omega)$ to vanish at $\omega = 0$ a decreasing density of
states in the renormalized electron spectrum is reasonable. In order to
resolve the observed structures in the renormalized electron spectrum we need
to examine larger systems.

\subsection{Static properties}
\label{subsec:statics}

\begin{figure}[ht]
\centering
\epsfig{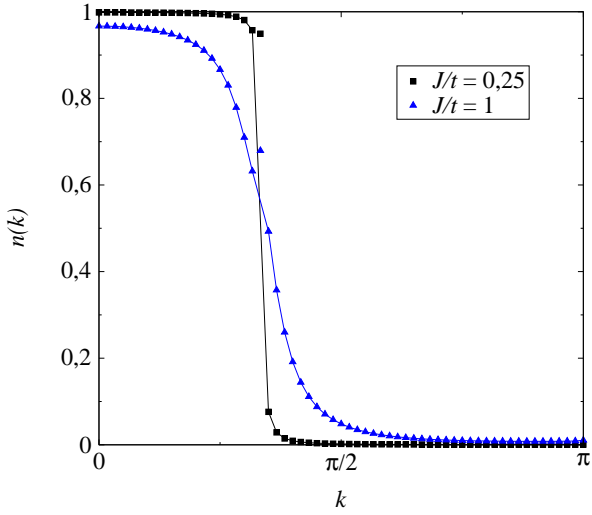}
\caption{Moment distribution function $n(k)$ at $n_c = 2/3$ and
different values $J/t$. The lines are drawn by ommiting the point at
$k = k^c_F$.}
\label{fig:nk}
\end{figure}

Let us now study the static correlation functions calculated in the last
section. The first quantity we want to consider is the momentum distribution
function $n(k)$ which is shown in Fig. \ref{fig:nk}. We obtain meaningful
results only for couplings up to $J/t \approx 1$. This signals a breakdown of
the flow equation treatment. In order to get better results for larger ratios
$J/t$ we need to go beyond the third order corrections in the flow
equations. Looking at $n(k)$ we notice that it is smeared out around
$k_F^c = n_c \pi / 2$. However, we can not decide whether these results
support the expected Luttinger liquid picture or not. The special behaviour of
$n(k)$ at $k = k_F^c$ may be due to the dominant peak structure of
$S(q)$. Since it is difficult to resolve the sharp peak of $S(q)$
appropriately for a finite system, we are not sure whether the artifact at $k
= k_F^c$ is due to the finite system size or the approximations. Nevertheless,
the shape of the momentum distribution function tends to support a small Fermi
surface, because there is no feature at $k =
\left(n_c+1\right)\pi/2$. Additionally, one may question if the effective
model (\ref{eq:eff_modell}) is capable of describing a large Fermi
surface. The system of conduction electrons within the effective model has a
Fermi momentum $n_c \pi/2$. Therefore a singularity in the momentum
distribution function is likely to be expected only at the point $k = k_F^c$.

\begin{figure}[ht]
\centering
\epsfig{figure=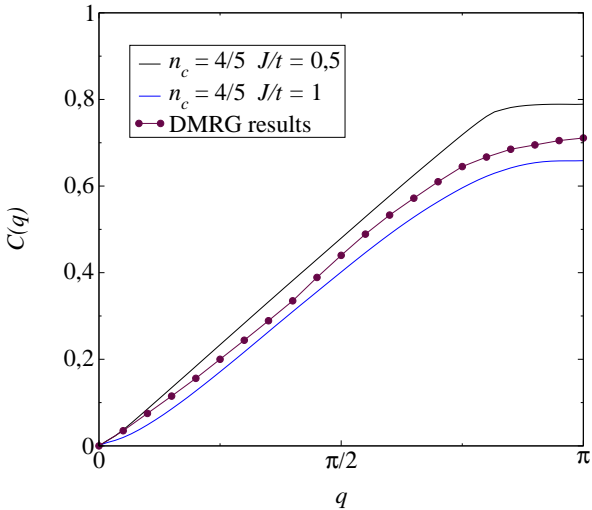,width=.9\columnwidth}
\caption{Charge correlation function $C(q)$ of the electrons for
different values $J/t$ and band fillings $n_c$. Circles are drawn for
comparison, taken from \cite{xavier_miranda} at $J/t = 1$ and $n_c = 4/5$.}
\label{fig:Cq}
\end{figure}

We can get further information from the charge correlation function
$C(q)$. The results are depicted in Fig. \ref{fig:Cq}. As we expect for small
couplings $J/t$ the function $C(q)$ takes the form of a noninteracting
one-dimensional electron gas with a kink at $q = 2 k_F^c = n_c
\pi$. Increasing $J/t$ leads to a cusp-like behaviour of $C(q)$ at $q = 2
k_F^c$. In addition, the slope at $q = 0$ drops with growing interaction
stregth $J/t$. The results displayed in Fig. \ref{fig:Cq} agree qualitatively
with the findings from numerical treatments \cite{tsunetsugu_rmp,
  moukouri_caron} in the examined parameter regime ($J/t \lesssim 1$). This
supports the Luttinger liquid picture of our description.

The charge correlation function gives us the possibility to derive the
parameter $K_\rho$ of the Luttinger liquid theory. This parameter is connected
to the slope of $C(q)$ at $q = 0$ via the relation
\cite{daul_noack}
\begin{equation}
K_\rho = \pi \, \frac{\partial C(q)}{\partial q} \, \bigg|_{q=0}.
\end{equation}

\begin{figure}[ht]
\centering
\epsfig{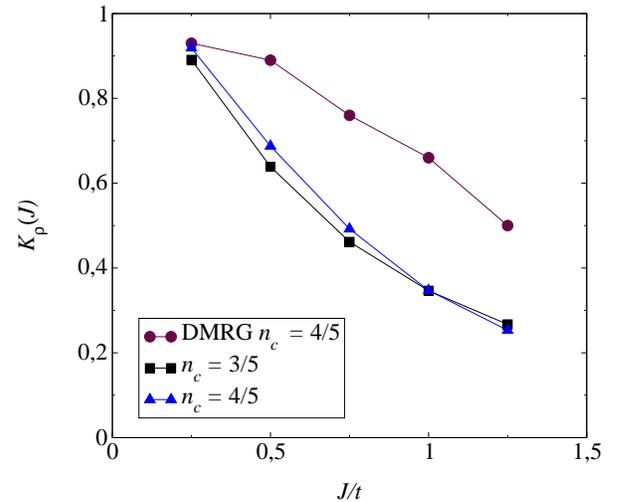}
\caption{Parameter $K_\rho$ as function of $J/t$ for two different band
  fillings $n_c$. Circles are drawn for comparison, data taken from
  \cite{xavier_miranda} for $n_c = 4/5$.}
\label{fig:K_r_J}
\end{figure}

The outcome is depicted in Fig. \ref{fig:K_r_J} as a function of the Kondo
coupling $J/t$. As we have already mentioned before the slope of $C(q)$ at $q
= 0$ decreases with growing coupling strength (up to the allowed value of $J/t
\lesssim 1$). For vanishing interaction strength $K_\rho \to 1$ corresponding
to a noninteracting electron gas. This can be understood from the equation for
$C(q)$ given in the appendix. Due to the flow equation
(\ref{eq:flussgleichung_gamma}) for the parameter $\gamma_{kq}$ of the
electron operator transformation all terms vanish which represent corrections
to the charge correlation function of independent electrons. Our findings are
in qualitative agreement with recent numerical results from DMRG calculations
\cite{xavier_miranda}. Xavier and Miranda find a minimum of $K_\rho(J)$ at
$J/t \approx 1.5$. Remember that our largest possible coupling is smaller than
$1.5$. Quantitatively our results are always considerably smaller than the
values found in ref. \cite{xavier_miranda}. Another work by Shibata \emph{et
  al.} \cite{shibata_jp} gives results for large $J/t$. In contrast to our
findings and to those of Xavier and Miranda \cite{xavier_miranda} these
authors expect $K_\rho \to 0$ in the limit $J/t \to 0$.

\begin{figure}[ht]
\centering
\epsfig{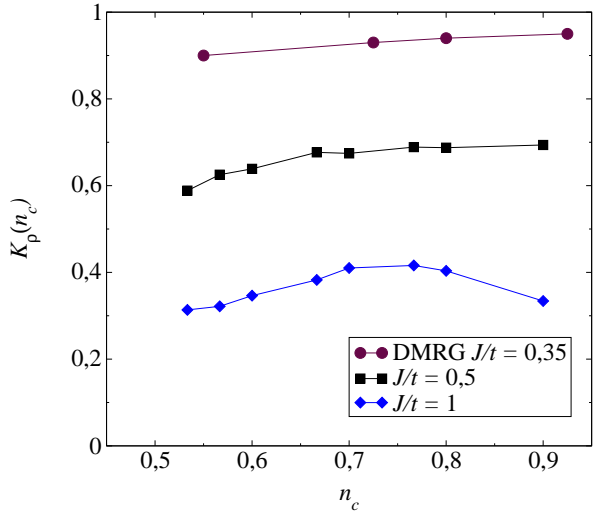}
\caption{Parameter $K_\rho$ as function of the band filling $n_c$ for
  different values $J/t$. Circles are drawn for comparison, data taken from
  \cite{xavier_miranda} for $J/t = 0.35$.}
\label{fig:K_r_nu}
\end{figure}

We also considered the dependence of $K_\rho$ on the band filling $n_c$. This
is depicted in Fig. \ref{fig:K_r_nu}. The lower possible value of the band
filling is $n_c = 1/2$ as we do not obtain a solution of the flow equations
below this value within the present approach. At small values $J/t$ we find a
monotonic decrease by lowering $n_c$. Again we find qualitative agreement with
Xavier and Miranda \cite{xavier_miranda}. As we already mentioned in the last
discussion our values for $K_\rho$ are considerably smaller compared to the
numerical data. For larger $J/t$ the behaviour deviates even qualitatively
from the numerical DMRG data. Whereas in ref. \cite{xavier_miranda} for all
values of $J/t$ a monotonic increase was obtained on increasing $n_c$, we find
a maximum in the function $K_\rho(n_c)$.

The magnetic properties of the one-dimensional KLM are significant for the
determination of the phase transition from the paramagnetic metallic phase
into the ferromagnetic phase. The spin correlation function for the conduction
electrons $S_{cc}(q)$ as well as for the local moments $S_{ff}(q)$ show a
characteristic increase of the ferromagnetic component $q = 0$ on approaching
the quantum phase transition.
\begin{figure}[ht]
\centering
\epsfig{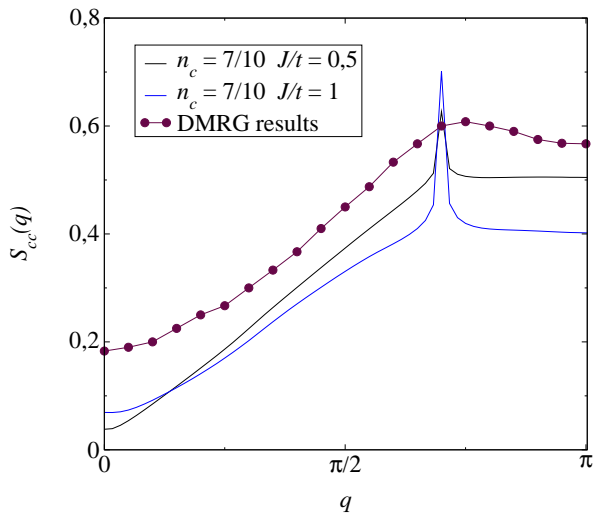}
\caption{Spin correlation function $S_{cc}(q)$ of the electrons for different
  values $J/t$ and band fillings $n_c$. Circles are drawn for comparison, data
  taken from \cite{moukouri_caron} for $J/t = 1$ and $n_c = 7/10$.}
\label{fig:Scc}
\end{figure}

The spin correlation function of the electrons $S_{cc}(q)$ is shown in
Fig. \ref{fig:Scc}. The strong peak at $q = 2 k_F^c$ results from the sharp
maximum of the spin correlation function $S(q)$ of the effective model. This
can easily be seen from the expression of $S_{cc}(q)$ given in the
appendix. Another characteristic is the finite weight of the ferromagnetic
component $q = 0$ which is directly connected with the occurence of the
quantum phase transition. On approaching the critical $J/t$ the maximum of
$S_{cc}(q)$ at $q = 2 k_F^c$ loses weight in favour of the ferromagnetic
component. This behaviour marks the phase transition
\cite{tsunetsugu_rmp}. As we already mentioned the present approach is
restricted to values of $J/t \lesssim 1$. These values are too small compared
to the value at the transition point which is $J/t \lesssim 2.5$ for $n_c =
2/3$ \cite{shibata_jp}. Nevertheless, we observe some tendency towards the
magnetic phase transition. As in the case of the charge correlation function
we compare our results with numerical data from \cite{moukouri_caron}. One can
clearly see the qualitative agreement between the two approaches, although our
findings tend to be smaller than the DMRG results. This is important if one
considers the points at $q = 0$. The increase of the ferromagnetic component,
which signals the tendency towards the quantum phase transition, turns out to
be comparably weak.
\begin{figure}[ht]
\centering
\epsfig{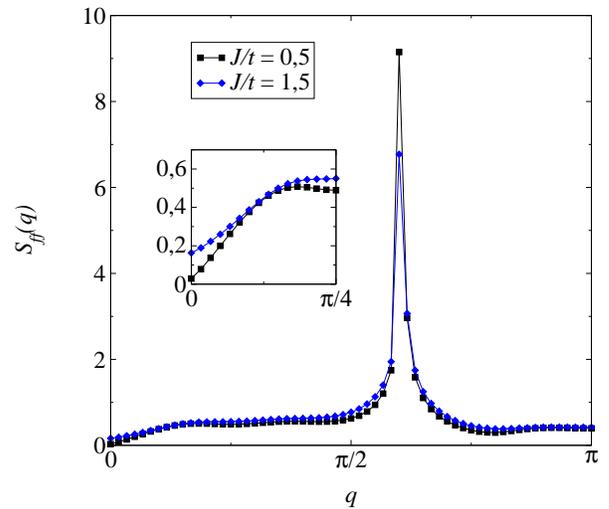}
\caption{Spin correlation function $S_{ff}(q)$ of the local moments at
$n_c = 3/5$ and different values $J/t$}
\label{fig:Sff}
\end{figure}

The situation we have just described is also characteristic for the spin
correlation function $S_{ff}(q)$ of the local moments which is drawn in
Fig. \ref{fig:Sff}. For small couplings we see that the corrections to the
spin correlation function of the effective model $S(q)$ are negligibly
small. Even for larger values of $J/t$ we find only small corrections. The
vicinity of the ferromagnetic component $q = 0$ is shown in the
inset. Nevertheless, the qualitative behaviour is once again in agreement with
numerical results \cite{tsunetsugu_rmp} though the values are somewhat
larger. Again, the ferromagnetic component gets an increasing weight while the
$q = 2 k_F^c$ component is suppressed.

\subsection{Dynamic properties}
\label{subsec:dynamics}

In the last section we have presented the results for static expectation
values and correlation functions. We have found that our results are in
qualitative agreement with numerical data for small couplings $J/t$. The flow
equation method sets us in the position to calculate not only static but also
dynamic correlation functions. Within our approach the dynamics of the KLM is
described in terms of the effective model (\ref{eq:eff_modell}). Since
$\tilde{\cal H}$ is blockdiagonal the dynamics for electrons and local spin
moments seperate. The SBMFT allows us, at least approximately, to characterise
the excitations of the spin system. The excitations of the KLM are determined
by a noninteracting Fermi gas (conduction electrons) and the Schwinger
bosons. In this section we shall add new aspects to the results obtained by
\cite{shibata_tsunetsugu}.
\begin{figure}[ht]
\centering
\epsfig{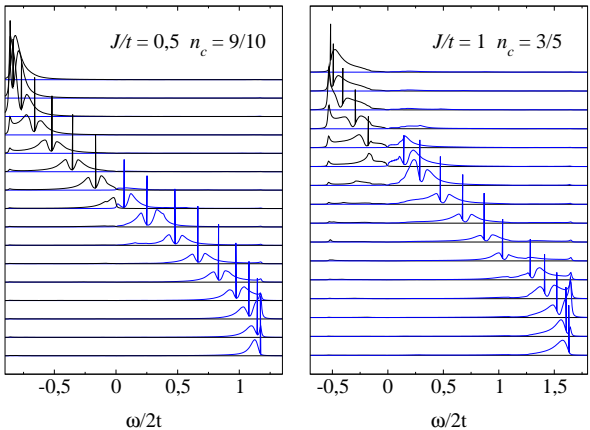}
\caption{Dynamic correlation functions $A_+(k,\omega)$ (black) and
$A_-(k,\omega)$ (blue) for $k = 0$ (up) to $k = \pi$ (down).}
\label{fig:dynamik_el}
\end{figure}

We start with the dynamic properties of the conduction electrons. The first
quantities we want to consider are the electronic spectral functions $A_\pm (k,
\omega)$ which can be measured in XPS and inverse XPS experiments. The outcome
is shown in Fig. \ref{fig:dynamik_el}. The energy is measured with respect to
the Fermi-energy of the conduction electrons $\varepsilon^c_F = n_c \pi/2$. As
we have already mentioned in the last section both functions consist of two
parts. A coherent quasiparticle-like contribution embodied by the finite peak
which has a weight $\tilde{\alpha}_k$. Its position is simply given by the
renormalized single-particle energies $\tilde{\varepsilon}_k$. The incoherent
background contains pairs of elementary excitations of the spin system of
$\tilde{\cal H}$. This follows directly from Eq. (\ref{eq:operator_e}) since
within the Schwinger boson approach for the effective spin system the
corrections to the spectral functions $A_\pm (k, \omega)$ are always connected
to the creation (annihilation) of pairs of bosons. The coupling to the
continuum of Schwinger boson excitations and the results for
$\tilde{\gamma}_{kq}$ give rise to the two maxima around the
quasiparticle-like peak.
\begin{figure}[ht]
\centering
\epsfig{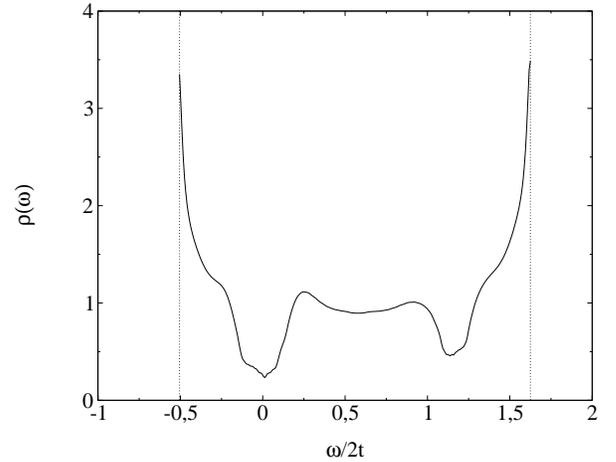}
\caption{Elektronic density of states $\rho(\omega)$ for $J/t = 1$
and $n_c = 3/5$.}
\label{fig:rho}
\end{figure}

The electronic density of states $\rho(\omega)$ is an important quantity which
shows a characteristic behaviour for Luttinger liquids. It is drawn in
Fig. \ref{fig:rho}. Again, the energy is measured with respect to
$\varepsilon^c_F = n_c \pi/2$. We find two minima, one in the vicinity of the
Fermi level of the conduction electrons, $\omega = 0$, the other above the
Fermi level. This behaviour follows from the pseudogap-like behaviour of the
renormalized single-particle energies $\tilde{\varepsilon}_k$. In contrast to
our findings DMRG studies from Shibata and Tsunetsugu
\cite{shibata_tsunetsugu} do not yield a minimum but rather a peak structure
just below $\omega = 0$ indicating the development of a pseudo gap. However,
their results were performed at finite temperatures. The Luttinger liquid
theory predicts a density of states following $\rho(\omega) \sim
|\omega|^\alpha$, $0 < \alpha < 1$ in the vicinity of $\omega = 0$. As can be
seen from Fig. \ref{fig:rho} there is no real vanishing of $\rho(\omega)$ at
$\omega = 0$. As we are dealing here with a finite system size we are not able
to resolve $\rho(\omega)$ near $\omega = 0$ and to verify the expected
behaviour.

Let us now turn to the magnetic properties. We want to present the results for
the dynamic spin structure factors of the electrons $S_{cc}(q, \omega)$ and
the local moments $S_{ff}(q, \omega)$. They describe the magnetic excitations
of the coupled system and can be measured by inelastic neutron scattering
experiments.

\begin{figure}[ht]
\centering
\epsfig{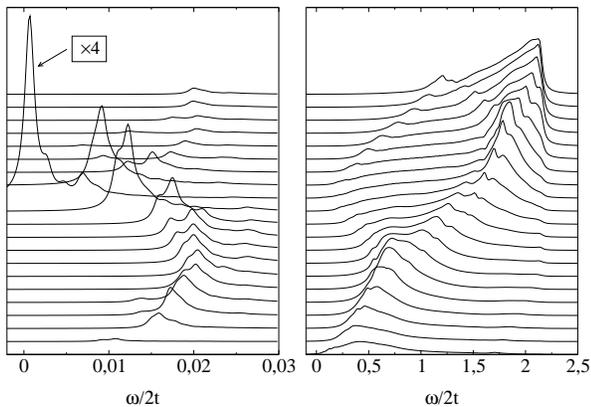}
\caption{Dynamic spin structure factor of the electrons
$S_{cc}(q,\omega)$ for $J/t = 1$ and $n_c = 3/5$, $q = 0$ (bottom) and
$q = \pi$ (top), Left: Low-energy part. Right: High-energy part.}
\label{fig:Sqw_el}
\end{figure}
We begin with the electronic dynamic spin structure factor $S_{cc}(q, \omega)$
which consists of a low- and a high-energy part. Both are discussed
seperately. The low energy sector, left panel of Fig. \ref{fig:Sqw_el}, is
characterised by the spin part of the effective model $\tilde{\cal{H}}_S$,
\emph{i.e.} the continuum of pair excitations of the Schwinger bosons. The
dominant contribution is therefore found at $q = 2 k_F^c$. It is multiplied by
a factor $1/4$ for a better comparison. We also see that there are regions
where no excitations are possible. Furthermore, the gap in the spectrum of the
elementary excitations $\tilde{\omega}_q$ leads to a gap in the low-energy
part of $S_{cc}(q, \omega)$. The high-energy sector of $S_{cc}(q, \omega)$ is
shown in the right panel of Fig. \ref{fig:Sqw_el}. The spectral weights are
about 10 times smaller compared to the weights of the low-energy part. The
main contribution arises from electronic particle-hole excitations of the
effective model. The specific form of this contribution shows therefore the
characteristics of a one-dimensional electron gas: a gapless excitation at $q
= 2 k_F^c$ and regions between $0 < q < 2 k_F^c$ where no excitations are
possible. In addition to the terms describing pure particle-hole excitations
there are also terms involving the elementary excitations $\tilde{\omega}_q$
of the effective spin system. They are responsible for the broadening of the
structures in the high energy sector of $S_{cc}(q, \omega)$.\\
The DMRG calculations of Shibata \emph{et al.} for $S_{cc}(\omega) = \int
\frac{dq}{2\pi} S_{cc}(q, \omega)$ showed a small peak at very low energies
and a larger double peak structure at higher energies
\cite{shibata_tsunetsugu}. We obtain a similar peak structure, but in contrast
to the results of \cite{shibata_tsunetsugu} the spectral weight of the low
energy part is much larger than the spectral weight of the high energy
part. This does not agree with the picture of an exhaustion of the electronic
low-energy spin degrees of freedom due to singlet formation described by
\cite{shibata_tsunetsugu}.

\begin{figure}[ht]
\centering
\epsfig{figure=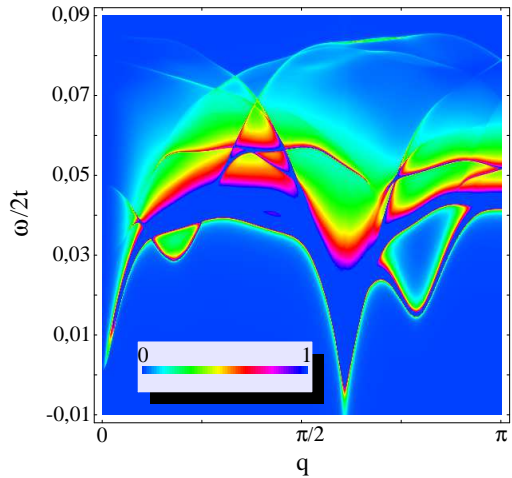, width=.8\columnwidth}
\caption{Low energy part of the dynamic spin structure factor
$S_{ff}(q,\omega)$ of the local moments for $J/t = 1.5$ and $n_c =
3/5$. The colour function is normalised to the maximum contribution.}
\label{fig:S_0_loc}
\end{figure}
Finally we want to discuss the magnetic excitations of the system of local
spin moments described by the dynamic spin structure factor $S_{ff}(q,
\omega)$. As in the case of the electronic spin structure factor $S_{cc}(q,
\omega)$ this function comprises a low- and a high-energy part. The first
one is again determined by the elementary excitations $\tilde{\omega}_q$ of
the spin part of the effective model $\tilde{{\cal H}}_S$. It is depicted in
Fig. \ref{fig:S_0_loc} and possesses the same features as the low-energy part
of $S_{cc}(q, \omega)$. From this picture we can clearly see the influence of
the low-energy spin excitations. The distinct structure at $q = 2 k_F^c$ gives
rise to the pronounced peak in the static spin correlation function
$S(q)$. Once again we point out that the energy scale of these excitations are
quite small compared with the effective band width of the electrons. The DMRG
results of Shibata \emph{et al.} for $S_{ff}(\omega) = \int \frac{dq}{2\pi}
S_{ff}(q, \omega)$ show a large peak structure at very small energies
\cite{shibata_tsunetsugu}. They assume that this is due to collective spin
excitations of the Luttinger-liquid. In our approach the low-energy peak is
the result of the continuum of elementary excitations of the effective spin
system, which we described in terms of Schwinger bosons.
 
\begin{figure}[ht]
\centering
\epsfig{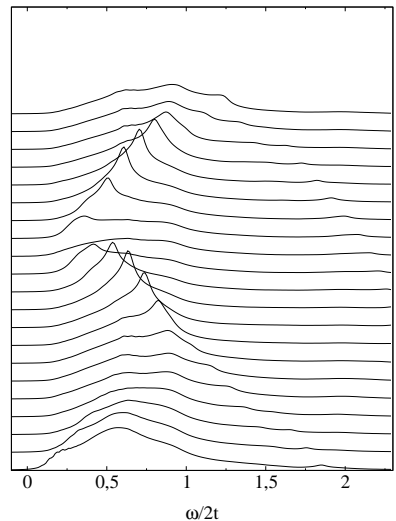}
\caption{High energy part of the dynamic spin structure factor
$S_{ff}(q,\omega)$ of the local moments at $J/t = 1.5$ and $n_c = 3/5$ for $q
  = 0$ (bottom) and $q = \pi$ (top).}
\label{fig:Sqw.loc}
\end{figure}
The high-energy part of $S_{ff}(q, \omega)$ is shown in
Fig. \ref{fig:Sqw.loc}. As in the case of $S_{cc}(q, \omega)$ it is mainly
determined by particle-hole excitations of the Fermi sea.  At larger couplings
$J/t$ the elementary spin excitations $\tilde{\omega}_q$ lead to the
broadening of the peak structure. Shibata \emph{et al.} obtain a second peak
in the high-energy sector of $S_{ff}(\omega)$ \cite{shibata_tsunetsugu}. Our
approach yields a similar structure in the local spin dynamics, although the
spectral weight of the high-energy part is much smaller than the spectral
weight of the low-energy part.\\
We further note that $S_{ff}(q, \omega)$ exhibits a finite gap. This is an
artifact and due to the approximations that we have made for the spin operator
transformation. By taking into account higher correction terms the transformed
spin operator of the local spin moment couples to the spin operator of the
conduction electrons. This gives rise to a gapless mode in $S_{ff}(q,
\omega)$.

\section{conclusion}
\label{sec:conclusion}

In summary, we used the method of continuous unitary transformations (flow
equation method) to examine the one-dimensional KLM. The renormalization
procedure was employed to integrate out the coupling between conduction
electrons and local spin operators. In that way we derived an effective
Hamiltonian which consists of an one-dimensional noninteracting electron gas
and a Heisenberg chain interacting via an RKKY-like coupling. In order to
treat the spin chain we used a Schwinger boson mean field theory
(SBMFT). Thereby we were able to calculate static and dynamic correlation
functions. The investigation of the electronic momentum distribution revealed
a small Fermi surface. We gave arguments, referring to the effective
model, why we were not able to obtain the large Fermi surface scenario in our
approach. Nevertheless, the static spin and charge correlation functions of
the electrons agreed qualitatively with numerical results. In addition we
obtained the parameter $K_\rho$ of the Luttinger liquid theory and found also
qualitative agreement with recent DMRG calculations. The present approach
was restricted to parameter regimes $J/t \lesssim 1$. Although the quantum
phase transition from the paramagnetic metallic into the ferromagnetic phase
takes place at larger values, we observed some tendency to a stronger
ferromagnetic component in the static spin correlation functions. The new
aspect of this work was the extension of calculations for dynamic properties
by means of the flow equation's method. We showed that the electronic spectral
functions comprised a coherent quasiparticle-like peak determined by the
renormalized electronic dispersion relation. The coupling to the low-energy
excitation of the effective spin model gave an incoherent background
comprising two maxima near the quasiparticle-like peaks. Finally, we also
computed the magnetic excitations of both the electrons and the local
spins. The corresponding spin structure factors always consisted of a
low-energy part, determined by the Schwinger boson pair excitations, and a
high-energy part, mostly determined by electronic particle-hole
excitations. The latter therefore showed the special features of the
one-dimensional Fermi surface. The electronic spin structure factor exhibited
a gapless mode at $q = 2 k_F^c$. Our results for the electronic spin dynamics
did not agree with the exhaustion picture described by
\cite{shibata_tsunetsugu}. The gapless mode at $q = 2 k_F^c$ should also be
seen in the spin structure factor of the local moments. There we argued that
further corrections in the spin operator transformation would lead to a
gapless mode.

\begin{acknowledgments}
The author would like to thank K. W. Becker, D. Efremov and K. Meyer
for helpful discussions and hints. This work was supported by the
Deutsche Forschungsgemeinschaft (DFG) through the research programme
SFB 463, Dresden.
\end{acknowledgments}

\appendix

\section{correlation functions in the flow equation approach} 

In this appendix we give the rather lengthy expressions for the static
and dynamic correlation functions omitted in the text. These are the
static charge correlation function $C(q)$ which reads
\begin{widetext}
\begin{align}
C(q) &= \frac{1}{N} \sum_{k k' \sigma \sigma'} \langle
c^\dagger_{k+q\sigma} c_{k\sigma} c^\dagger_{k'-q\sigma'}
c_{k'\sigma'} \rangle
\allowdisplaybreaks \notag \\
&= \frac{2}{N} \sum_k \, \tilde{\alpha}^2_k \, \tilde{\alpha}^2_{k+q}
\, n_{k+q} (1-n_k)
\allowdisplaybreaks \notag \\
&+ \frac{4}{N^2} \sum_{kp} \, \tilde{\alpha}_k \,
\tilde{\alpha}_{k+q} \, \tilde{\gamma}_{p,k+q} \,
\tilde{\gamma}_{p-q,k} \, S(k-p-q) \, n_{k+q} (1-n_k)
+ \frac{2}{N^2} \sum_{kp} \, \tilde{\alpha}^2_k \,
\tilde{\gamma}^2_{k-q,p} \, S(k-p-q) \, n_k (1-n_p)
\allowdisplaybreaks \notag \\
&+ \frac{2}{N^2} \sum_{kp} \, \tilde{\alpha}^2_k \,
\tilde{\gamma}^2_{k+q,p} \, S(k-p+q) \, n_p (1-n_k)
+ \frac{4}{N^2} \sum_{kp} \, \tilde{\alpha}_k \,
\tilde{\alpha}_{p} \, \tilde{\gamma}_{p+q,k} \,
\tilde{\gamma}_{k-q,p} \, S(k-p-q) \, n_k ( 1 - n_p )
\label{eq:C_q}
\end{align}
The dynamic spin structure factor $S_{cc}(q,\omega)$ of the electrons
takes the form
\begin{align}
S_{cc}(q,\omega) &= \int_{-\infty}^{\infty} dt \,
\langle {\bf s}_q (t) \cdot {\bf s}_{-q} \rangle \, \text{e}^{i \omega t}
\allowdisplaybreaks \notag \\
&=\frac{3}{2 N} \sum_k \, \tilde{\alpha}^2_k \,
\tilde{\alpha}^2_{k+q} \, n_{k+q} (1-n_k) \,
\delta(\omega-\tilde{\varepsilon}_k+\tilde{\varepsilon}_{k+q})
\allowdisplaybreaks \notag \\
&- \frac{1}{N^2} \sum_{kp} \, \tilde{\alpha}_k \, \tilde{\alpha}_{k+q}
\, \tilde{\gamma}_{p-q,k} \, \tilde{\gamma}_{p,k+q} \, S(k-p-q) \,
n_{k+q} (1 - n_k) \,
\delta(\omega-\tilde{\varepsilon}_k+\tilde{\varepsilon}_{k+q})
\allowdisplaybreaks \notag \\
&- \frac{1}{2 N^3} \sum_{kpp'} \, \tilde{\alpha}_k \, \tilde{\alpha}_{p}
\, \tilde{\gamma}_{p+q,k} \, \tilde{\gamma}_{k-q,p} \, (u_{p'}
v_{p'+k-p-q}-v_{p'} u_{p'+k-p-q})^2 \, n_k ( 1 - n_p ) 
\delta(\omega-\tilde{\omega}_{p'}-\tilde{\omega}_{p'+k-p-q}-
\tilde{\varepsilon}_k+\tilde{\varepsilon}_p)
\allowdisplaybreaks \notag \\
&+ \frac{3}{4 N^3} \sum_{kpp'} \tilde{\alpha}^2_k \,
\tilde{\gamma}^2_{k+q,p} \, (u_{p'} v_{p'+k-p+q}-v_{p'} u_{p'+k-p+q})^2
\, n_p (1 - n_k) 
\delta(\omega-\tilde{\omega}_{p'}-\tilde{\omega}_{p'+k-p+q}-
\tilde{\varepsilon}_p+\tilde{\varepsilon}_k)
\allowdisplaybreaks \notag \\
&+ \frac{3}{4 N^3} \sum_{kpp'} \tilde{\alpha}^2_k \,
\tilde{\gamma}^2_{k-q,p} \, (u_{p'} v_{p'+k-p-q}-v_{p'} u_{p'+k-p-q})^2
\, n_k (1 - n_p) 
\delta(\omega-\tilde{\omega}_{p'}-\tilde{\omega}_{p'+k-p-q}-
\tilde{\varepsilon}_k+\tilde{\varepsilon}_p)
\allowdisplaybreaks \notag \\
&+ \frac{1}{2 N^3} \sum_{kpp'} \, \tilde{\alpha}_k \, \tilde{\alpha}_{p}
\, (\tilde{\gamma}_{k+q,k} \, \tilde{\gamma}_{p-q,p} +
\tilde{\gamma}_{k-q,k} \, \tilde{\gamma}_{p+q,p})
n_k \, n_p \, (u_{p'} v_{p'+q}-v_{p'} u_{p'+q})^2 \,
\delta(\omega-\tilde{\omega}_{p'}-\tilde{\omega}_{p'+q})
\allowdisplaybreaks \notag \\
&+ \frac{1}{2 N^3} \sum_{kpp'} \, \tilde{\alpha}_k \, \tilde{\alpha}_{p}
\, (\tilde{\gamma}_{k-q,k} \, \tilde{\gamma}_{p-q,p} +
\tilde{\gamma}_{k+q,k} \, \tilde{\gamma}_{p+q,p})
n_k \, n_p \, (u_{p'} v_{p'+q}-v_{p'} u_{p'+q})^2 \,
\delta(\omega-\tilde{\omega}_{p'}-\tilde{\omega}_{p'+q})
\label{eq:Scc_qw}
\end{align}
Here, it can clearly be seen that the second and third line involves
only particle hole excitations of the Fermi sea of the effective
model. The last two lines represent the low energy sector of
$S_{cc}(q, \omega)$ as they include only pair excitations of Schwinger
bosons. On integrating over the energy $\omega$ one obtains the
expression for the static spin correlation function $S_{cc}(q)$.
\end{widetext}


\begin{thebibliography}{50}
%
%
\bibitem{fulde} Peter Fulde {\itshape Electron Correlation in
         Molecules and Solids}, Third Enlarged Edition,
         Springer-Verlag, Berlin Heidelberg, 1995
\bibitem{hewson} A. C. Hewson {\itshape The Kondo Problem to Heavy
        Fermions}, Cambridge University Press, Cambridge, 1993
\bibitem{troyer_wuertz} M. Troyer, D. W\"urtz,
        Phys. Rev. B {\bf 47}, 2886 (1993)
\bibitem{tsunetsugu_rmp} Hirokazu Tsunetsugu, Manfred Sigrist, Kazuo
        Ueda, Rev. Mod. Phys. {\bf 69}, 809 (1997) \emph{and
        references therein}
\bibitem{tsunetsugu_prb} Hirokazu Tsunetsugu, Kazuo Ueda, Manfred Sigrist,
        Phys. Rev. B {\bf 47}, 8345 (1993)
\bibitem{shibata_jp} Naokazu Shibata, Kazuo Ueda, J. Phys.:
        Condens. Matter {\bf 11}, R1 (1999)
\bibitem{xavier_1} J.C. Xavier, E. Novais, E. Miranda,
        Phys. Rev. B {\bf 65}, 214406 (2002)
\bibitem{moukouri_caron} S. Moukouri, L. Caron,
        Phys. Rev. B {\bf 52}, R15723 (1995)
\bibitem{caprara_rosengren} S. Caprara, A. Rosengren,
        Europhys. Lett. {\bf 39}, 55 (1997)
\bibitem{shibata_tsunetsugu} N. Shibata and H. Tsunetsugu
        J. Phys. Soc. Jpn. {\bf 68}, 3138 (1999)
\bibitem{yu_white} Clare C. Yu, Steven R. White,
        Phys. Rev. Lett. {\bf 71}, 3866 (1993)
\bibitem{honner_gulacsi} Graeme Honner, Miklos Gulacsi,
        Phys. Rev. Lett. {\bf 78}, 2180 (1997); Graeme Honner, Miklos
        Gulacsi, Phys. Rev. B {\bf 58}, 2662 (1998)
\bibitem{mcculloch} I. P. McCulloch, A. Juozapavicius, A. Rosengren,
        M. Gulacsi, Phys. Rev. B {\bf 65}, 052410 (2002)
\bibitem{pivovarov_si} Eugene Pivovarov, Qimiao Si,
        Phys. Rev. B {\bf 69}, 115104 (2004)
\bibitem{doniach} S. Doniach, Physica B {\bf 91}, 231 (1977)
\bibitem{luttinger} J. M. Luttinger, Phys. Rev. {\bf 119}, 1153 (1960)
\bibitem{voit} J. Voit, Rep. Prog. Phys. {\bf 57}, 977 (1994)
\bibitem{schrieffer_wolff}  J. R. Schrieffer, P. A. Wolff, 
        Phys. Rev. {\bf 149}, 491 (1966)
\bibitem{shibata_new} N. Shibata \emph{et al.} cond-mat/0503476
\bibitem{wegner} Franz Wegner, Ann. Physik {\bf 3}, 77 (1994)
\bibitem{glazek_wilson} S. G\l azek, K. G. Wilson, 
        Phys. Rev. D {\bf 48}, 5863 (1994)
\bibitem{stein} J. Stein, Eur. Phys. J. B {\bf 12}, 5 (1999)
%
%
\bibitem{electron_phonon} P. Lenz, F. Wegner, \emph{Nucl. Phys. B}
        {\bf 482} 693 (1996);
	M. Ragwitz, F. Wegner, Eur. Phys. J. B {\bf 8}, 9 (1999);
        A. Mielke, Ann. Physik {\bf 6}, 215 (1997);
	A. Mielke \emph{Europhys. Lett.} {\bf 40} ({\bf 2}),
	pp. 195-200 (1997);
\bibitem{1d_systeme} C. Heidbrink, G. Uhrig, 
        Phys. Rev. Lett. {\bf 88}, 146401 (2002); C. Heidbrink,
        G. Uhrig, Eur. Phys. J. B {\bf 30}, 443 (2002); S. Kehrein,
        Phys. Rev. Lett. {\bf 83}, 4914 (1999)
\bibitem{spin_boson} G. Uhrig, Phys. Rev. B {\bf 57}, R14004 (1998);
        C. Raas, U. L\"ow, G. Uhrig, Phys. Rev. B {\bf 65}, 144438 (2002);
	S. Kehrein, A. Mielke, Ann. Physik {\bf 6} 90 (1997);
	S. Kehrein, A. Mielke, P. Neu, Z. Phys. B {\bf 99} 269 (1996)
%
%
\bibitem{arovas_auerbach} Daniel P. Arovas, Assa Auerbach,
        Phys. Rev. B {\bf 38} 316 (1988); Assa Auerbach, Daniel
        P. Arovas, Phys. Rev. Lett. {\bf 61} 617 (1988)
\bibitem{trumper_manuel_gazza_ceccatto} A. E. Trumper, L. O. Manuel,
        C. J. Gazza, H. A. Ceccatto,
        Phys. Rev. Lett. {\bf 78}, 2216 (1997) 
\bibitem{ceccatto_gazza_trumper} H. A. Ceccatto, C. J. Gazza,
        A. E. Trumper, Phys. Rev. B {\bf 47}, 12 329 (1993)
%
%
\bibitem{daul_noack} S. Daul, R.M. Noack, Phys. Rev. B {\bf 58}, 2635
        (1998)
\bibitem{xavier_miranda} J.C. Xavier, E. Miranda, Phys. Rev. B {\bf 70},
        075110 (2004)
\bibitem{sachdev} S. Sachdev, Phys. Rev. B {\bf
        45}, 12 377 (1992)
%
%
\end{thebibliography}
\end{document}